\documentclass{article}

    \PassOptionsToPackage{numbers, compress}{natbib}

\usepackage[final]{neurips_2024}

\usepackage[utf8]{inputenc} 
\usepackage[T1]{fontenc}    
\usepackage{hyperref}       
\usepackage{url}            
\usepackage{booktabs}       
\usepackage{amsfonts}       
\usepackage{nicefrac}       
\usepackage{microtype}      
\usepackage{xcolor}         
\usepackage{wrapfig}
\usepackage{graphicx}
\usepackage{bm}
\usepackage{amssymb}
\usepackage{multirow}
\usepackage{makecell}
\usepackage{bbding}
\usepackage{pifont}

\newcommand{\tabincell}[2]{\begin{tabular}{@{}#1@{}}#2\end{tabular}}

\title{\textsc{Frieren}: Efficient Video-to-Audio Generation Network with Rectified Flow Matching}

\author{%
  Yongqi~Wang\thanks{Equal contribution.}, Wenxiang~Guo\footnotemark[1], Rongjie~Huang, Jiawei~Huang, Zehan~Wang, 
 \\ \textbf{Fuming~You, Ruiqi~Li, Zhou~Zhao\thanks{Corresponding author.}} \\ 
  Zhejiang University \\
  \texttt{cyanbox@zju.edu.cn} \\
}

\begin{document}

\maketitle

\begin{abstract}

Video-to-audio (V2A) generation aims to synthesize content-matching audio from silent video, and it remains challenging to build V2A models with high generation quality, efficiency, and visual-audio temporal synchrony. 
We propose \textsc{Frieren}, a V2A model based on rectified flow matching. \textsc{Frieren} regresses the conditional transport vector field from noise to spectrogram latent with straight paths and conducts sampling by solving ODE, outperforming autoregressive and score-based models in terms of audio quality. By employing a non-autoregressive vector field estimator based on a feed-forward transformer and channel-level cross-modal feature fusion with strong temporal alignment, our model generates audio that is highly synchronized with the input video. Furthermore, through reflow and one-step distillation with guided vector field, our model can generate decent audio in a few, or even only one sampling step. Experiments indicate that \textsc{Frieren} achieves state-of-the-art performance in both generation quality and temporal alignment on VGGSound, with alignment accuracy reaching 97.22\%, and 6.2\% improvement in inception score over the strong diffusion-based baseline. Audio samples and code are available at \url{http://frieren-v2a.github.io}.
\end{abstract}

\section{Introduction}
\label{sec:intro}

Recent advancements in deep generative models have significantly enhanced the quality and diversity of AI-generated content, including text \cite{radford2018improving}, images \cite{rombach2022high, saharia2022photorealistic, betker2023improving}, videos \cite{singer2022make, liu2024sora} and audios \cite{liu2023audioldm, liu2023audioldm2}. Among various content-generation tasks, video-to-audio (V2A) generation aims to synthesize semantically relevant and temporally aligned audio from video frames. Due to its immense potential for application in film dubbing, game development, YouTube content creation and other areas, the task of V2A has attracted widespread attention.

A widely applicable V2A solution is expected to have outstanding performance in the following aspects: \textbf{1) audio quality}: the generated audio should have good perceptual quality, which is the fundamental requirement of the audio generation task; \textbf{2) temporal alignment}: the generated audio should not only match the content but also align temporally with the video frames. This has a significant impact on user experience due to keen human perception of audio-visual information; and \textbf{3) generation efficiency}: the model should be efficient in terms of generation speed and resource utilization, which affects its practicality for large-scale and high-throughput applications.

Currently, considerable methods have been proposed for this task, including GAN-based models \cite{chen2020generating, ghose2022foleygan}, transformer-based autoregressive models \cite{iashin2021taming, sheffer2023hear}, and a recent latent-diffusion-based model, Diff-Foley \cite{luo2024diff}. However, these methods have not yet achieved a balanced and satisfactory performance across the above aspects. 1) For audio quality, early GAN-based models suffer from poor quality and lack practicality. Autoregressive and diffusion models make improvements in generation quality, but still leave room for further advancement. 2) For temporal alignment, autoregressive models lack the ability to align the generated audio with the video explicitly. And due to the difficulty of learning audio-visual alignment with the cross-attention-based conditional mechanism solely, Diff-Foley relies on additional classifier guidance to achieve good synchrony, which not only increases the model complexity but also leads to instability when reducing sampling steps. 3) For generation efficiency, autoregressive models suffer from high inference latency, while Diff-Foley requires considerable sampling steps to achieve good generation quality due to the curved sampling trajectories of diffusion models, increasing the temporal overhead in inference. In a nutshell, existing methods still leave significant room for improvement in performance.

In this paper, We introduce another generative modeling approach, namely rectified flow matching~\cite{liu2022flow}, into the V2A task. This method regresses the conditional transport vector field between noise and data distributions with as straight trajectories as possible, and conducts sampling by solving the corresponding ordinary differential equation (ODE). With simpler formulations, our rectified-flow-based model achieves higher audio quality and diversity. To improve temporal alignment, we adopt a non-autoregressive vector field estimator network with a feed-forward transformer with no temporal-dimension downsampling, thereby preserving temporal resolution. We also employ a channel-level cross-modal feature fusion mechanism for conditioning, leveraging the inherent alignment of audio-visual data and achieving strong alignment. These designs lead to high synchrony between generated audio and input video while upholding model simplicity. Moreover, through integrating reflow and one-step distillation techniques, our model can generate decent audio with a few, or even only one sampling step, significantly improving generation efficiency.

We name our model \textsc{Frieren} for e\textbf{f}ficient video-to-audio gene\textbf{r}at\textbf{i}on n\textbf{e}twork with \textbf{re}ctfied flow matchi\textbf{n}g. Experiments indicate that \textsc{Frieren} outperforms strong baselines in terms of audio quality, generation efficiency, and temporal alignment on VGGSound \cite{chen2020vggsound}, achieving a 6.2\% improvement in inception score (IS) and a generation speed 7.3$\times$ that of Diff-Foley, as well as temporal alignment accuracy of up to 97.22\% in 25 steps. Additionally, \textsc{Frieren} combining reflow and distillation achieves alignment accuracy of up to 97.85\% with just one step, with a 9.3$\times$ acceleration compared to 25-step sampling, further boosting generation efficiency.
\section{Related works}

\subsection{Video-to-audio generation}

Video-to-audio (V2A) generation aims to synthesize audio of which content matches the visual information of a video clip. RegNet \cite{chen2020generating} designs a time-dependent visual encoder to extract appearance and motion features, which are then fed to a GAN for audio generation. FoleyGAN \cite{ghose2022foleygan} also utilizes GAN for audio generation from visual features, together with a predicted action category as the conditional input. SpecVQGAN \cite{iashin2021taming} takes RGB and optical flow of videos and uses a transformer to generate indices of a spectrogram VQVAE autoregressively. Im2Wav \cite{sheffer2023hear} adopts two transformers for different temporal resolutions and takes CLIP \cite{radford2021learning} features as the condition to generate VQVAE indices. Du et al. \cite{du2023conditional} mimics the real-world foley methodology and introduces an additional reference audio as the condition. Diff-Foley \cite{luo2024diff} designs an audio-visual contrastive feature and adopts a latent diffusion to predict spectrogram latents, achieving decent audio quality and inference speed.

In addition to training a whole model from scratch, some works integrate off-the-shelf audio generation models with modality mappers or multimodal encoders with joint embedding space for conditioning. V2A-Mapper \cite{wang2024v2a} uses a lightweight mapper to transfer CLIP embeddings of videos to CLAP \cite{wu2023large} embeddings as the condition for audio generation. Xing et al. \cite{xing2024seeing} utilize an ImageBind\cite{girdhar2023imagebind}-based latent aligner for conditional guidance in audio generation. Despite the existence of plentiful works on V2A, there is still a large room left for improvement in quality, synchrony, and efficiency.

\subsection{Flow matching generative models}

Flow matching \cite{lipman2022flow} models the vector field of transport probability path from noise to data samples. Compared to score-based models like DDPM \cite{ho2020denoising}, flow matching achieves more stable and robust training together with superior performance. Specifically, rectified flow matching \cite{liu2022flow} learns the transport ODE to follow the straight paths connecting the noise and data points as much as possible, reducing the transport cost, and achieving fewer sampling steps with the reflow technique. This modeling paradigm has demonstrated excellent performance in accelerating image generation \cite{liu2023instaflow, esser2024scaling}.

In the area of audio generation, Voicebox \cite{le2024voicebox} builds a large-scale multi-task speech generation model based on flow matching. Its successor, Audiobox \cite{vyas2023audiobox}, extends the flow-matching-based model to a unified audio generation model with natural language prompt guidance. Matcha-tts \cite{mehta2024matcha} trains an encoder-decoder TTS model with optimal-transport conditional flow matching. VoiceFlow \cite{guo2024voiceflow} introduces rectified flow matching into TTS, achieving speech generation with fewer inference steps. However, for the task of V2A, there has been no exploration into utilizing flow matching models to enhance generation quality or inference efficiency.

\section{Method}

\subsection{Preliminary: rectified flow matching}

We first introduce the basic principles of rectified flow matching (RFM) \cite{liu2022flow} that we build our model upon. Conditional generation problems like V2A can be viewed as a conditional mapping from a noise distribution $\bm{x}_0 \sim p_0(\bm{x})$ to a data distribution $\bm{x}_1 \sim p_1(\bm{x})$. This mapping can be further taken as a time-dependent changing process of probability density (a.k.a. flow), determined by the ODE: 
\begin{equation}
    \mathrm{d}\bm{x} = \bm{u}(\bm{x}, t | \bm{c}) \mathrm{d}t , t \in [0, 1],
\end{equation}
where $t$ represents the time position, $\bm{x}$ is a point in the probability density space at time $t$, $\bm{u}$ is the value of the transport vector field (i.e., the gradient of the probability w.r.t $t$) at $\bm{x}$, and $\bm{c}$ is the condition. In our case, the condition $\bm{c}$ is the visual features from the video frames, while the data $\bm{x}_1$ is the compressed mel-spectrogram latent of the corresponding audio from a pre-trained autoencoder. The fundamental principle of flow matching generative model is to use a neural network $\theta$ to regress the vector field $\bm{u}$ with the flow matching objective:
\begin{equation}
    \mathcal{L}_{\mathrm{FM}}(\theta)=\mathbb{E}_{t, p_t(\bm{x})}\left\|\bm{v}(\bm{x}, t|\bm{c};\theta)-\bm{u}(\bm{x}, t|c)\right\|^2,
    \label{eq:fm_objective}
\end{equation}

\begin{wrapfigure}{r}{0.36\textwidth}
    \vspace{-1em}
    \includegraphics[width=0.35\textwidth]{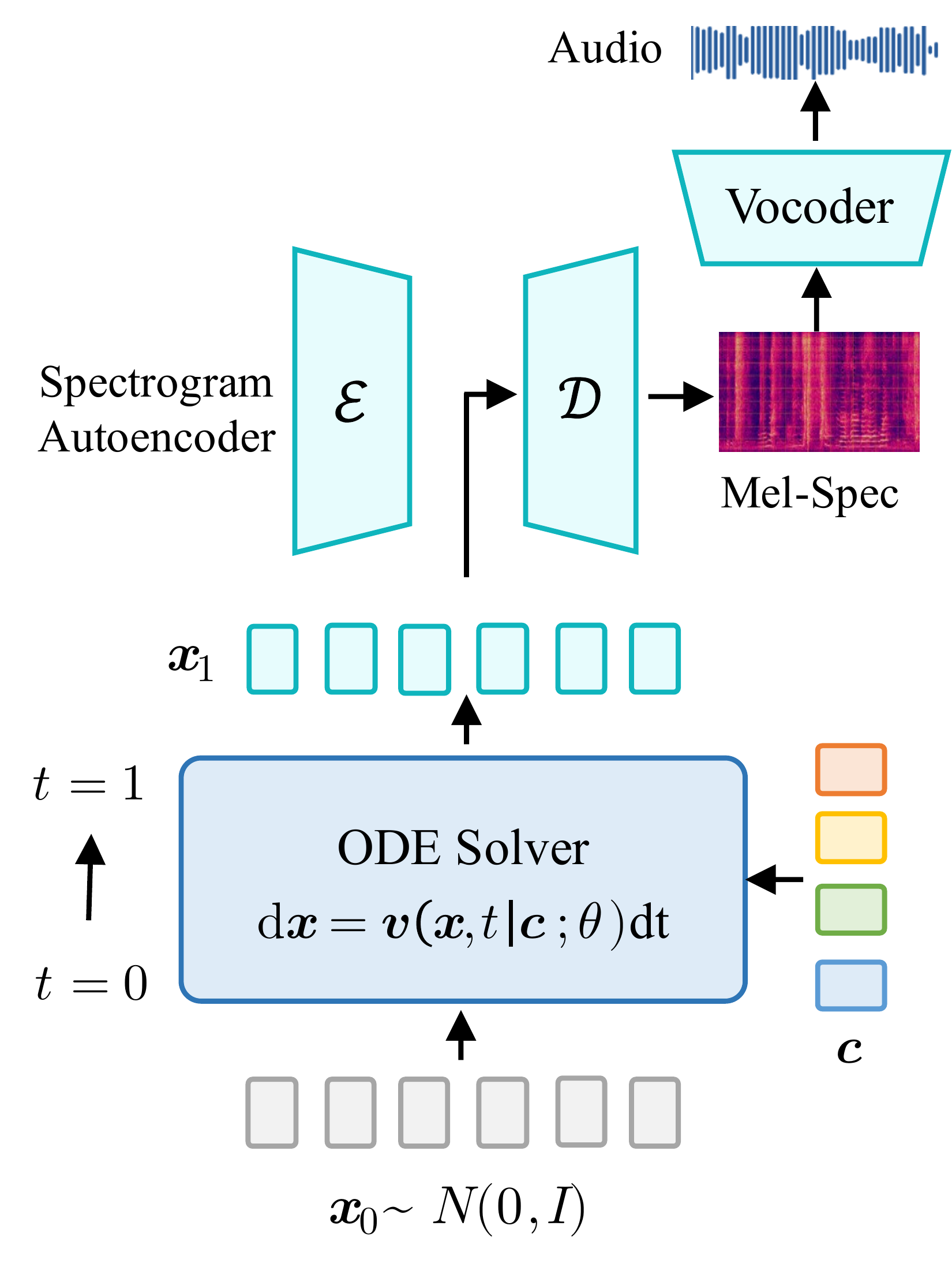}
    \caption{Illustration of the sampling process of our rectified-flow based V2A architecture.}
    \label{fig:ode}
\end{wrapfigure}

where $p_t(\bm{x})$ is the distribution of $\bm{x}$ at timestep $t$. However, due to a lack of prior knowledge of target distribution $p_1(\bm{x})$ and the forms of $p_t$ and $\bm{u}$, it is intractable to directly compute $\bm{u}(\bm{x}, t|c)$. As an alternative, conditional flow matching objective, which is proven in \cite{lipman2022flow} to have identical gradient as eq.~\ref{eq:fm_objective} w.r.t $\theta$, is used for regression:
\begin{equation}
    \mathcal{L}_{\mathrm{CFM}}(\theta)=\mathbb{E}_{t, p_1(\bm{x_1}),p_t(\bm{x}|\bm{x}_1)}\left\|\bm{v}(\bm{x}, t|\bm{c};\theta)-\bm{u}(\bm{x}, t|\bm{x}_1,c)\right\|^2.
    \label{eq:cfm_objective}
\end{equation}

Through designing specific probabilistic paths that enable efficient sampling from $p_t(\bm{x}|\bm{x}_1)$ and computing of $\bm{u}(\bm{x}, t|\bm{x}_1,c)$, we achieve an unbiased estimation of $\bm{u}(\bm{x}, t|,c)$ with the CFM objective \ref{eq:cfm_objective}. Specifically, rectified flow matching attempts to establish straight paths between noise and data, aiming to facilitate sampling with larger step sizes and fewer steps. Given a noise-data pair $(\bm{x}_0, \bm{x}_1)$, $\bm{x}$ is located at $(1-t)\bm{x}_0 + t\bm{x}_1$ at timestep $t$, with the vector field being $\bm{u}(\bm{x}, t|\bm{x}_1,c)=\bm{x}_1 - \bm{x}_0$, pointing from the noise point to the data point. Hence, for each training step of the vector field estimator, we simply sample the data point $\bm{x}_1$ and noise point $\bm{x}_0$ from $p_1(\bm{x})$ and $p_0(\bm{x})$, respectively, and optimize the network with the rectified flow matching (RFM) loss
\begin{equation}
    \left\|\bm{v}(\bm{x}, t|\bm{c};\theta)-(\bm{x}_1 - \bm{x}_0)\right\|^2 .
    \label{eq:rfm_objective}
\end{equation}

Once the vector estimator network finishes training, we can adopt various solvers to approximate the solution of the ODE $\mathrm{d}\bm{x} = \bm{v}(\bm{x}, t | \bm{c}; \theta)$ at discretized time steps for sampling. A simple and commonly used ODE solver is the Euler method:
\begin{equation}
    \bm{x}_{t+\epsilon} = \bm{x} + \epsilon \bm{v}(\bm{x}, t | \bm{c}; \theta)
    \label{eq:euler}
\end{equation}
where $\epsilon$ is the step size. The sampled latent is fed to the decoder of the spectrogram autoencoder for spectrogram reconstruction, and the result is further used to reconstruct the audio waveform with a vocoder. Figure~\ref{fig:ode} provides a simple demonstration of the model's sampling process.

\subsection{Model architecture}

\begin{figure}[t]
  \centering
   \vspace{-1em}
  \hspace*{-0.5em}\includegraphics[width=\textwidth]{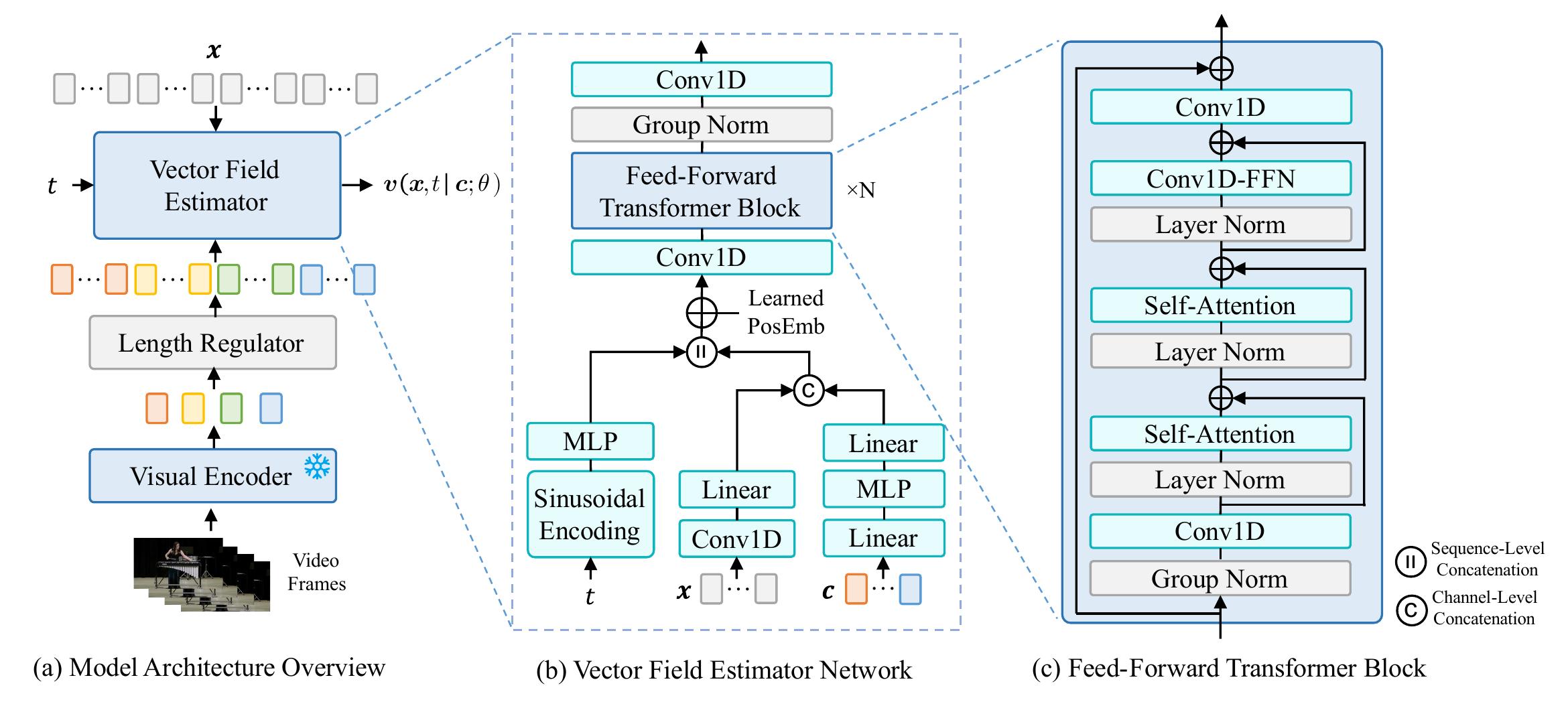}
  \caption{Illustration of model architecture of \textsc{Frieren} at different levels.}
  \vspace{-1em}

  \label{fig:arch}
\end{figure}

\paragraph{Model overview} We illustrate the model architecture of \textsc{Frieren} at different levels in Figure \ref{fig:arch}. As shown in Figure \ref{fig:arch}(a), we first utilize a pre-trained visual encoder with frozen parameters to extract a frame-level feature sequence from the video. Usually, the video frame rate is lower than the temporal length per second of the spectrogram latent. To align the visual feature sequence with the mel latent at the temporal dimension for the cross-modal feature fusion mentioned below, we adopt a length regulator, which simply duplicates each item in the feature sequence by the ratio of the latent length per second and the video frame rate for regulation. The regulated feature sequence is then fed to the vector field estimator as the condition, together with $\bm{x}$ and $t$, to get the vector field prediction $\bm{v}$.

\paragraph{Visual and audio representations}
Various audio-aligned visual representations \cite{girdhar2023imagebind, luo2024diff, huang2024mavil, wang2023connecting, wang2023extending, wangfreebind, wang2024omnibind} can potentially be applied to video-to-audio generation, and we conduct experiments with two types of visual representations. For a fair comparison with Diff-Foley \cite{luo2024diff}, we mainly utilize the CAVP feature proposed in \cite{luo2024diff}, which is a visual-audio contrastive feature considering both content and temporal alignment. Meanwhile, to investigate the impact of visual feature characteristics on model performance, we also attempt the visual feature from MAViL \footnote{Implementation of MAViL is from av-superb \cite{tseng2024av}: \url{https://github.com/roger-tseng/av-superb}} \cite{huang2024mavil}, which is an advanced self-supervised visual-audio representation learner that employs both masked-reconstruction and contrastive learning, and exhibits formidable performance in audio-visual understanding (See section \ref{sec:abl_feature} for comparison). For audio representation, we follow a previous text-to-audio work \cite{huang2023make} to train a mel-spectrogram VAE with 1D convolution over the temporal dimension. Details of the VAE are provided in appendix \ref{appendix:param}.

\paragraph{Vector field estimator} Figure \ref{fig:arch}(b) demonstrates the structure of the vector field estimator, which is composed of a feed-forward transformer and some auxiliary layers. The regularized visual feature $\bm{c}$ and the point $\bm{x}$ on the transport path are first processed by stacks of shallow layers separately, with output dimensions being both half of the transformer hidden dimension, and are then concatenated along the channel dimension to realize cross-modal feature fusion. This simple mechanism leverages the inherent alignment within the video and audio, achieving enforced alignment without relying on learning-based mechanisms such as attention. As a result, the generated audio and input video sequences exhibit excellent temporal alignment. After appending the time step embedding to the beginning, the sequence is added with a learnable positional embedding and is then fed into the feed-forward transformer. The structure of the transformer block is illustrated in Figure \ref{fig:arch} (c), the design of which is derived from the spatial transformer in latent diffusion \cite{rombach2022high}, with the 2D convolution layers replaced by 1D ones. The feed-forward transformer does not involve temporal downsampling, thus preserving the resolution of the temporal dimension and further ensuring the preservation of alignment. The output of the stacked transformer blocks is then passed through a normalization layer and a 1D convolution layer to finally obtain the prediction of the vector field.

\subsection{Re-weighting RFM objective with logit-normal coefficient}

The original RFM objective samples uniformly over time span $[0,1]$. However, for modeling the vector field, positions in the middle of the transport path (equivalent to time steps in the middle of [0,~1]) present greater difficulty, as these positions are distant from both noise and data distributions. On the other hand, positions near the boundaries of the time span typically lie close to corresponding noise or data points, and their vector field direction tends to align with the lines connecting these points and the centroid of the distribution on the opposite side, and therefore relatively easy to regress. Upon this insight, we introduce time-based re-weighting to the original RFM objective, allocating more weight to intermediate time steps to achieve better modeling effectiveness. This is equivalent to increasing the sampling frequency of intermediate time steps. 
In practice, logit-normal weighting coefficients have been proven \cite{esser2024scaling} to yield promising results, with the formula being
\begin{equation}
    w(t) = \frac{1}{\sqrt{2\pi}}\frac{1}{t(1-t)}\exp{\left(-\frac{(\ln t - \ln (1-t))^2}{2}\right)} .
    \label{eq:reweight}
\end{equation}

We re-weight the RFM objective with this weighting function to replace the original objective and observe in our experiment that this re-weighting helps to slightly improve audio quality and temporal alignment at the cost of a marginal decrease in audio diversity.

\subsection{Classifier-free guidance}

Similar to diffusion-based models, we observe that classifier-free guidance (CFG) is highly important for generating audio that semantically matches and temporally aligns with the video. During training, we randomly replace the condition sequence $\bm{c}$ with a zero tensor with a probability of 0.2, and during sampling, we modify the vector field using the formula
\begin{equation}
    \bm{v}_{\mathrm{CFG}}(\bm{x}, t|\bm{c};\theta) = \gamma \bm{v}(\bm{x}, t|\bm{c};\theta) + (1-\gamma) \bm{v}(\bm{x}, t|\varnothing ;\theta) ,
\end{equation}
where $\gamma$ is the guidance scale trading off the sample diversity and generation quality, and $\bm{v}_{\mathrm{CFG}}$ degenerates into the original vector field $\bm{v}$ when $\gamma=1$. We set $\gamma$ to 4.5 in our major experiments.

\subsection{Reflow and one-step distillation with guided vector field}

In this section, we introduce two techniques we adopt for reducing sampling steps. The first one is reflow, which is a crucial component of the rectified flow paradigm \cite{liu2022flow, liu2023instaflow}. Training the estimator network with objective \ref{eq:rfm_objective} for once is insufficient to construct straight enough transport paths, and an extra reflow procedure is needed to strengthen the transport trajectories without altering the marginal distribution learned by the model, enabling sampling with larger step sizes and fewer steps. Given a model $\theta$ trained with RFM objective, the reflow procedure applies $\theta$ to conduct sampling over the entire training dataset to obtain sampled data $\hat{\bm{x}}_1$ and save the corresponding input noise $\bm{x}_0'$, finally obtaining triplets $(\bm{x}_0', \hat{\bm{x}}_1, \bm{c})$. The noise-data pair $(\bm{x}_0, \bm{x}_1)$ in the RFM objective \ref{eq:rfm_objective} is replaced by $(\bm{x}_0', \hat{\bm{x}}_1)$ for a secondary training of $\theta$. This process can be repeated multiple times to obtain straighter trajectories with diminishing marginal effects. We conduct reflow for once as it is sufficient for achieving straight enough trajectories.

While many rectified-flow-based models regress the same velocity field $\bm{v}$ during both the initial training and the reflow process, we observe that when incorporating CFG, conducting sampling and reflow with the original vector field $\bm{v}$ is ineffective in straightening the sampling trajectories with the guided vector field $\bm{v}_{\mathrm{CFG}}$. Therefore, we use $\bm{v}_{\mathrm{CFG}}$ for generating $\hat{\bm{x}}_1$ and as the target of regression in reflow. The reflow objective can be written as:
\begin{equation}
    \mathcal{L}_{\mathrm{reflow}}(\theta')=\mathbb{E}_{t, p(\bm{x}_0', \hat{\bm{x}}_1|\bm{c}),p_t(\bm{x}|\bm{x}_0', \hat{\bm{x}}_1)}\left\|\bm{v}_{\mathrm{CFG}}(\bm{x}, t|\bm{c};\theta')-(\hat{\bm{x}}_1 - \bm{x}_0')\right\|^2
\end{equation}
with same weighting function as eq. \ref{eq:reweight}.

Upon the model $\theta'$ obtained from reflow, we further conduct one-step distillation \cite{liu2022flow, liu2023instaflow} to enhance the single-step generation performance of the model. As a type of self-distillation, this procedure tries to reduce the error between the single-step sampling result $\bm{x}_0' + \bm{v}_{\mathrm{CFG}}(\bm{x}_0', t|\bm{c};\theta)$ and the multi-step sampling result $\hat{\bm{x}}_1$. The objective function can be written as:
\begin{equation}
    \mathcal{L}_{\mathrm{distill}}(\theta'')=\mathbb{E}_{t, p(\bm{x}_0', \hat{\bm{x}}_1|\bm{c}),p_t(\bm{x}|\bm{x}_0', \hat{\bm{x}}_1)}\left\|\bm{x}_0' + \bm{v}_{\mathrm{CFG}}(\bm{x}_0', t|\bm{c};\theta'')-\hat{\bm{x}}_1\right\|^2
    \label{eq:distill}
\end{equation}

Formally, the distillation objective \ref{eq:distill} can be viewed as a reflow objective with the sampling timestep fixed at $t=0$. We observe in the experiment that due to a limited number of sampling steps in reflow data generation, the model may experience a decrease in sampling quality after the reflow process. Therefore, we opt to use the same training data used in reflow for distillation, rather than re-sampling the training data with the reflow model, which is based on the theoretical basis that reflow does not alter the marginal distribution modeled by the estimator.

\section{Experiments}

\begin{table}[t]
    \caption{Results of V2A models on VGGSound dataset. R+F and RN50 denote the RGB+Flow and ResNet50 versions of SpecVQGAN, and CG denotes classifier guidance in Diff-Foley.}
    \label{tab:res1}
    \footnotesize
    \centering
    \resizebox{\textwidth}{!}{
    \begin{tabular}{lccccccccc}
    \toprule
    Model & FD$\downarrow$ & IS$\uparrow$ & KL$\downarrow$ & FAD$\downarrow$ & KID$(10^{-3}) \downarrow$ & Acc(\%) $\uparrow$ & MOS-Q$\uparrow$  & MOS-A$\uparrow$ \\ 
    \midrule
    SpecVQGAN (R+F) & 31.69 & 5.23 & 3.37 & 5.42 & 8.53 & 61.83 & 3.30 $\pm$ 0.06 & 2.35 $\pm$ 0.05 \\
    SpecVQGAN (RN50) & 32.52 & 5.21 & 3.41 & 5.39 & 9.00 & 56.92 & 3.25 $\pm$ 0.07 & 2.17 $\pm$ 0.05 \\
    Im2Wav & 14.98 & 7.20 & \textbf{2.57} & 5.49  & 3.35 & 56.70 & 3.39 $\pm$ 0.06 & 2.29 $\pm$ 0.06 \\
    Diff-Foley (CG \ding{51}) & 23.94 & 11.11 & 3.38 & 4.72 & 9.58 & 95.03 & 3.57 $\pm$ 0.08 & 3.74 $\pm$ 0.07  \\
    Diff-Foley (CG \ding{55}) & 24.97 & 11.69 & 3.23 & 7.10 & 10.32 & 92.53 & 3.64 $\pm$ 0.07 & 3.59 $\pm$ 0.06  \\
    LDM  & 11.79 & 10.09 &  2.86 & 1.77 & \textbf{2.26} & 95.33 & 3.72 $\pm$ 0.05 & 3.79 $\pm$ 0.07  \\
    \textsc{Frieren}  & 12.26 & 12.42 & 2.73 & \textbf{1.32} & 2.49 & \textbf{97.22} & 3.78 $\pm$ 0.06 & \textbf{3.90 $\pm$ 0.05} \\
    \textsc{Frieren} (Dopri5) & \textbf{11.64} & \textbf{12.76} & 2.75 & 1.37 & 2.39 & 96.87 & \textbf{3.81 $\pm$ 0.06} & 3.85 $\pm$ 0.06 \\
    \bottomrule 
    \end{tabular}
    }

\end{table}

\subsection{Experiment setup}
\label{subsec:exp_setup}

\paragraph{Dataset and pre-processing}

Following most previous works, we take VGGSound \cite{chen2020vggsound} as the benchmark, which consists of 200k+ 10-second video clips from YouTube spanning 309 categories. Excluding videos already removed from YouTube, we follow the original train and test splits of VGGSound, the sizes of which are about 182.6k and 15.3k. We downsample the audios to 16kHz and transform them to mel-spectrogram with 80 bins and a hop size of 256. We follow \cite{luo2024diff} to downsample the videos to 4 FPS. Data samples are truncated to 8-second clips for training and inference.

\paragraph{Model configuration}

The transformer of the vector field estimator mainly used in the experiments has 4 layers and a hidden dimension of 576. Each model is trained with 2 NVIDIA RTX-4090 GPUs. We train the estimator for 1.3M steps for the first training, and 600k and 500k steps for reflow and distillation, with the learning rate being 5e-5 for all stages. For waveform generation, we train a BigVGAN \cite{lee2022bigvgan} vocoder on AudioSet \cite{gemmeke2017audio}. Details of model parameters are provided in appendix \ref{appendix:param}.

\paragraph{Metrics}

We combine objective and subjective metrics to evaluate model performance over audio quality, diversity, and temporal alignment. For objective evaluation, we calculate Frechet distance (FD), inception score (IS), Kullback–Leibler divergence (KL), Frechet audio distance (FAD), kernel inception distance (KID), and alignment accuracy (Acc). We utilize audio evaluation tools provided by AudioLDM \cite{liu2023audioldm}, which are widely used in audio generation tasks, as well as the alignment classifier provided in \cite{luo2024diff}. For metrics with reference like FAD, we duplicate the reference audio samples in the test set for 10 times as we generate 10 samples for each data item. For subjective evaluation, we conduct crowd-sourced human evaluations with 1-5 Likert scales and report mean-opinion-scores (MOS) over audio quality (MOS-Q) and content alignment (MOS-A) with 95\% confidence intervals (CI). We sample 10 audios for each test video for evaluation. Details of subjective evaluation are provided in appendix \ref{appendix:eval}.

\paragraph{Baseline models}

We adopt three advanced V2A models as baselines, including: 1) SpecVQGAN \cite{iashin2021taming}, a transformer-based autoregressive model generating spectrogram VQVAE indices from visual features; 2) Im2Wav \cite{sheffer2023hear}, a hierarchical autoregressive V2A model predicting audio VQVAE indices conditioned on CLIP features; and 3) Diff-Foley \cite{luo2024diff}, a strong latent-diffusion-based V2A model. For SpecVQGAN, we evaluate two versions using RGB+Flow and ResNet features as input visual conditions. For Diff-Foley, we evaluate its performance with and without classifier guidance to examine the impact of its complex external alignment mechanism. To better validate the superiority of our rectified flow model, we also train a diffusion model sharing the same architecture as \textsc{Frieren} but has a different prediction target, labeled as LDM in the following tables. For diffusion models, we use DPM-Solver \cite{lu2022dpm} for sampling. For our rectified flow model, we use the Euler method \ref{eq:euler} in most cases without further specification. We also explore the more advanced Dormand–Prince method (Dopri5) \cite{dormand1980family} method for higher generation quality.

\subsection{Results and analysis}

\paragraph{Video-to-audio generation results}

The results of different models are illustrated in table \ref{tab:res1}. We sample with diffusion models and \textsc{Frieren} with 25 steps, and report the result of \textsc{Frieren} without reflow and distillation, which shows the best overall performance with a high number of sampling steps. It can be seen that \textsc{Frieren} significantly outperforms other models in IS, FAD, and alignment accuracy, with the values reaching up to 12.42, 1.32, and 97.22\%, together with high subjective scores of 3.78 and 3.90 on quality and alignment. For FD, KL, and KID, the scores of \textsc{Frieren} are also very close to the best values among other models. When we employ the higher-order Dopri5 ODE solver, \textsc{Frieren} achieves further improvements in FD and IS, attaining best values of 11.64 and 12.76, respectively, while maintaining stable performance in other objective metrics, at the cost of slower sampling speed. This indicates the effectiveness of our approach. Generally, the performance of \textsc{Frieren} surpasses that of the LDM, demonstrating the superiority of rectified flow matching over the score-based paradigm of diffusion. Additionally, both \textsc{Frieren} and LDM outperform Diff-Foley in temporal alignment, proving that our architecture design achieves strong temporal alignment without the need for complex mechanisms, and can produce audio that is highly synchronized with visual input. Additionally, our model also has an advantage in sampling time, with details provided in appendix \ref{appendix:time}.

\begin{table}[t]
    \caption{Results of \textsc{Frieren} and Diff-Foley under different sampling steps. CG denotes classifier guidance, R denotes reflow and D denotes one-step distillation.}
    \label{tab:res2}
    \scriptsize
    \centering
    \begin{tabular}{lcccccccccc}
    \toprule
    Model & Steps & FD$\downarrow$ & IS$\uparrow$ & KL$\downarrow$ & FAD$\downarrow$ & KID$(10^{-3}) \downarrow$ & Acc(\%) $\uparrow$ & MOS-Q$\uparrow$ & MOS-A$\uparrow$ \\ 
    \midrule
    Diff-Foley (CG \ding{51}) & \multirow{5}{*}{\tabincell{c}{1}} & 82.61 & 2.31 & 4.44 & 13.64 & 43.96 & 31.60 & 1.28 $\pm$ 0.04 & 1.35 $\pm$ 0.03 \\
    Diff-Foley (CG \ding{55}) & & 86.97 & 1.86 & 4.17 & 14.66 & 39.73 & 37.02 & 1.17 $\pm$ 0.03 & 1.63 $\pm$ 0.04 \\
    \textsc{Frieren} (R \ding{55}, D \ding{55}) &  & 70.48 & 2.95 & 4.21 & 13.07 & 26.99 &  43.18 & 2.12 $\pm$ 0.04 & 1.71 $\pm$ 0.04 \\
    \textsc{Frieren} (R \ding{51}, D \ding{55}) &  & 18.61 & 6.63 & 2.60 & 3.13 & 3.49 & 94.96 & 3.32 $\pm$ 0.07 & 3.74 $\pm$ 0.06 \\
    \textsc{Frieren} (R \ding{51}, D \ding{51}) & & \textbf{17.58} & \textbf{8.66} & \textbf{2.56} & \textbf{1.85} & \textbf{2.91} & \textbf{97.85} & \textbf{3.48 $\pm$ 0.06} & \textbf{3.93 $\pm$ 0.05}  \\
    \midrule
    Diff-Foley (CG \ding{51}) & \multirow{4}{*}{\tabincell{c}{5}} & 60.99 & 3.42 & 3.62 & 9.61 & 3.60 & 73.30 & 2.66 $\pm$ 0.07 & 2.98 $\pm$ 0.07 \\
    Diff-Foley (CG \ding{55}) &  & 51.52 & 5.14 & 3.45 & 10.96 & 2.66 & 91.30 & 3.03 $\pm$ 0.08 & 3.56 $\pm$ 0.07 \\
    \textsc{Frieren} (R \ding{55}, D \ding{55}) &  & 28.78 & 6.69 & 3.02 & 4.34 & 8.56 & 87.69 & 3.30 $\pm$ 0.07 & 3.37 $\pm$ 0.08 \\
    \textsc{Frieren} (R \ding{51}, D \ding{55}) &  & \textbf{14.65} & \textbf{8.28} & \textbf{2.60} & \textbf{2.11} & \textbf{2.28} & \textbf{96.82} & \textbf{3.43 $\pm$ 0.06} & \textbf{3.83 $\pm$ 0.06} \\
    \midrule
    Diff-Foley (CG \ding{51}) & \multirow{4}{*}{\tabincell{c}{25}} & 23.94 & 11.11 & 3.28 & 4.72 & 9.58 & 95.03 & 3.57 $\pm$ 0.08 & 3.74 $\pm$ 0.07 \\
    Diff-Foley (CG \ding{55}) &  & 24.97 & 11.69 & 3.23 & 7.10 & 10.32 & 92.53 & 3.64 $\pm$ 0.07 & 3.59 $\pm$ 0.06  \\
    \textsc{Frieren} (R \ding{55}, D \ding{55}) & & 12.26 & \textbf{12.42} & 2.73 & \textbf{1.32} & 2.49 &  97.22 & \textbf{3.78 $\pm$ 0.06} & \textbf{3.90 $\pm$ 0.05} \\
    \textsc{Frieren} (R \ding{51}, D \ding{55}) &  & 13.39 & 9.79 & \textbf{2.64} & 1.66 & 2.01 & \textbf{97.36} & 3.61 $\pm$ 0.07  & 3.88 $\pm$ 0.05 \\
    \bottomrule 
    \end{tabular}
\end{table}

\begin{figure}[t]
  \centering
  \includegraphics[width=\textwidth]{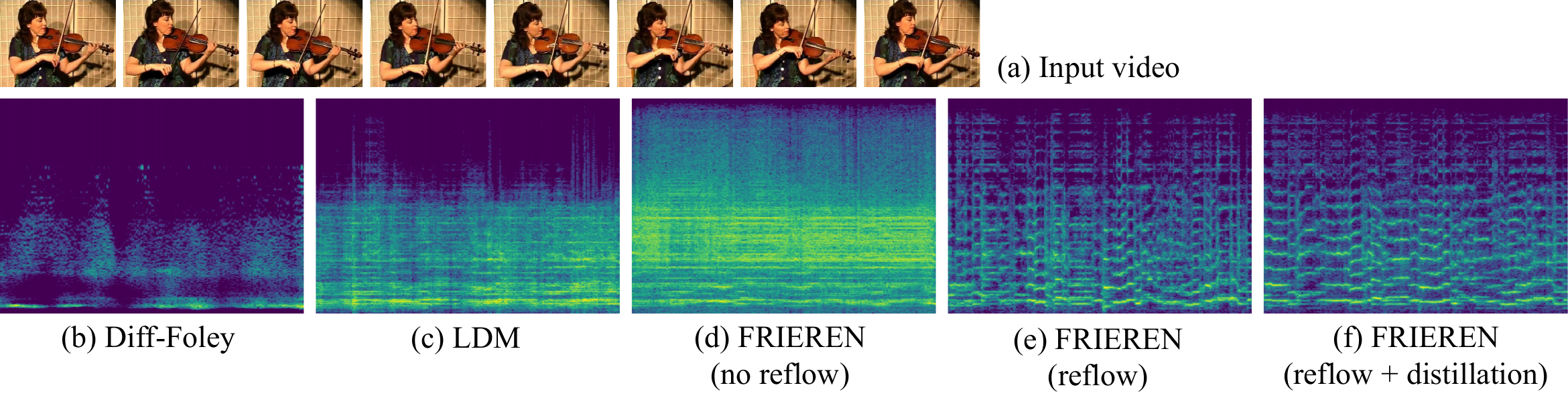}  
  \vspace{-2em}
  \caption{One-step generation results of different models. (a): The content of the input video is a woman playing the violin. (b): Diff-Foley generates meaningless audio with one step. (c, d): LDM and \textsc{Frieren} without reflow generate highly noisy audio. (e, f): reflow enables \textsc{Frieren} to generate meaningful audio in one step, and distillation further improves the one-step generation quality.}
  \vspace{-1em}
  \label{fig:1_step}
\end{figure}

\paragraph{Few and single step generation results}

\begin{wrapfigure}{r}{0.6\textwidth}
    \vspace{-1em}
    \includegraphics[width=0.6\textwidth]{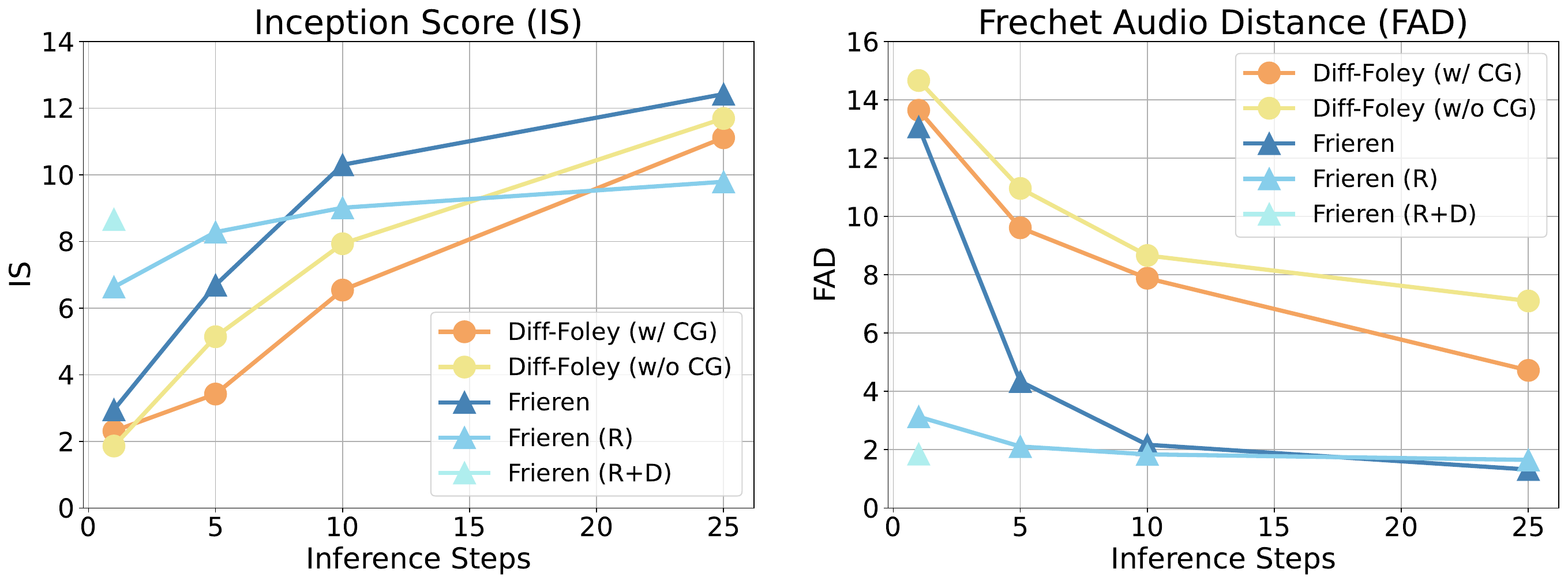}
    \caption{IS and FAD of the models with different steps.}
    \label{fig:fewstep}
    \vspace{-3mm}
\end{wrapfigure}

We further demonstrate the results of Diff-Foley and \textsc{Frieren} on reduced sampling steps in table \ref{tab:res2} to illustrate the impact of reflow and one-step distillation, together with trend graphs of IS and FAD in figure \ref{fig:fewstep} for intuitive presentation. The data for reflow are generated with the Euler method for 25 steps. We observe an obvious drop in performance of Diff-Foley as well as \textsc{Frieren} without reflow when sampling with as few as 5 steps, and their scores become extremely poor when we further reduce the step number to 1. Figure \ref{fig:1_step} (b) (c) and (d) illustrate that the audio generated by these models as well as LDM degrades into unacceptably noisy or meaningless audio within one step. This is due to the convoluted nature of the sampling trajectories of these models, which disables them from sampling with large step sizes and few steps. We also notice that when sampling with 5 steps, using additional classifier guidance deteriorates the audio quality and synchrony of Diff-Foley, where alignment accuracy and IS drop by 18.0\% and 1.72 respectively, while FD, KL, and KID increase by 9.47, 0.17, and 0.94 $\times 10^{-3}$. This indicates the lack of robustness of the complex alignment mechanism that Diff-Foley relies on. 

In contrast, \textsc{Frieren} with reflow achieves an alignment accuracy of up to 96.82\% in just 5 steps, with significant advantages in quality, diversity, and subjective metrics. Additionally, it maintains an accuracy of 94.96\% in single-step generation, as well as decent quality and diversity. This proves that reflow functions significantly in straightening the sampling trajectories, enabling the rectified flow model to generate decent audio with a small number of sampling steps. Furthermore, single-step distillation following reflow further improves the model performance with one step, with alignment accuracy reaching up to 97.85\%, and KL, FAD, and KID being close to the 25-step results of \textsc{Frieren} trained once, with differences of 0.17, 0.53 and 0.42$\times 10^{-3}$. It also achieves high MOS-Q and MOS-A of 3.48 and 3.93. 

Figure \ref{fig:1_step} (e) and (f) show that results from \textsc{Frieren} with reflow and reflow+distillation have distinguishable spectrograms, with the latter showing higher quality and sharper edges. This fully demonstrates that the combination of reflow and one-step distillation endows our model with strong single-step generation capabilities, significantly enhancing the efficiency on the V2A task. Notice that reflow brings in some quality degradation in sampling with 25 steps. We speculate that this is because the limited number of sampling steps restricts the data quality when generating data for reflow, resulting in a shift in the marginal distribution learned by the model. This cumulative error might be mitigated by increasing the number of sampling steps during reflow data generation.

\begin{table}[t]
    \caption{Ablation results on different model size of vector field estimator network.} 
    \label{tab:size}
    \small
    \centering
    \resizebox{\textwidth}{!}{
    \begin{tabular}{lcccccccccc}
    \toprule
    Model Size & FD$\downarrow$ & IS$\uparrow$ & KL$\downarrow$ & FAD$\downarrow$ & KID$(10^{-3}) \downarrow$ & Acc(\%) $\uparrow$ & MOS-Q$\uparrow$ & MOS-A$\uparrow$ \\ 
    \midrule
    Small (70.90 M) & 13.02 & 12.16 & 2.78 & 1.50 & 2.79 & 96.04 & 3.71 $\pm$ 0.07 & 3.83 $\pm$ 0.06 \\
    Base (158.88 M) & 12.26 & 12.42 & 2.73 & 1.32 & 2.49 & 97.22 & 3.78 $\pm$ 0.06 & 3.90 $\pm$ 0.05  \\
    Large (421.12 M) & 12.20 & 12.29 & 2.76 & 1.36 & 2.97 & 95.16  & 3.78 $\pm$ 0.07 & 3.80 $\pm$ 0.06 \\
    \bottomrule 
    \end{tabular}
    }
\end{table}

\subsection{Ablation study}

\subsubsection{Model size of vector field estimator}

We adjust the number of parameters of the vector field estimator and evaluate the model performance at different scales. We label the major model as ``base'', and obtain ``small'' and ``large'' models by decreasing and increasing the hidden dimension and / or the number of transformer layers, respectively. The parameter counts of the estimator and results are presented in table \ref{tab:size}.

We observe that when the model parameters are reduced to 71M, performance declines across all metrics, where FD, KL, FAD, and KID increase by 0.76, 0.05, 0.18, and 0.3$\times 10^{-3}$, and IS, alignment accuracy, MOS-Q and MOS-A drop by 0.26, 1.18\%, 0.07 and 0.07, respectively. However, when the parameter number increases to 421M, there is a performance degradation across multiple metrics, with KL, FAD, and KID increasing by 0.03, 0.04, and 0.48$\times 10^{-3}$, and IS, alignment acc declining by 0.13 and 2.06\%. We speculate that this anomalous phenomenon may be due to the convergence difficulty for the larger model under similar training steps, or the redundant model capacity tends to cause overfitting on a relatively small dataset like VGGSound, deteriorating the model's generalization performance. 
In summary, we achieve relatively balanced model performance with the parameter of the estimator being around 160M. Details of model parameters are provided in appendix \ref{appendix:param}.

\subsubsection{Visual feature characteristics}
\label{sec:abl_feature}

In table \ref{tab:abl_feature}, we compare the results of \textsc{Frieren} using two different types of visual features from CAVP and MAViL. Intuitively, the MAViL feature should be more robust and contain richer audio-related semantic information, as it utilizes masked-reconstruction together with inter-modal and intra-modal contrastive learning, in contrast to CAVP trained solely with inter-modal contrastive learning. On the other hand, however, due to MAViL's convolutional downsampling in the temporal dimension, its feature sequence has a lower effective FPS of 2 with the same 4 FPS video input as CAVP. The results in the table indicate that the model with MAViL feature excels in audio diversity, with differences of FD, KL, and FAD being 0.18, 0.24, and 0.06. Meanwhile, it exhibits a 7.05\% decrease in alignment accuracy and a 0.25 decrease in IS. This result yields two insights for V2A tasks: 1) at relatively low frame rates, the frame rate of features, rather than content, is more likely to become the bottleneck for audio quality and visual-audio synchrony; 2) compared to high video frame rates, the semantic information and robustness of visual features are more crucial for the diversity of generated audio.

\begin{table}[t]
    \caption{Results on different types visual features.}
    \label{tab:abl_feature}
    \small
    \centering
    \resizebox{\textwidth}{!}{
    \begin{tabular}{lcccccccccc}
    \toprule
    Type & Feat. FPS & FD$\downarrow$ & IS$\uparrow$ & KL$\downarrow$ & FAD$\downarrow$ & KID$(10^{-3}) \downarrow$ &  Acc(\%) $\uparrow$ & MOS-Q$\uparrow$ & MOS-A$\uparrow$ \\ 
    \midrule

    CAVP \cite{luo2024diff} & 4 & 12.26 & 12.42 & 2.73 & 1.32 & 2.49 &  97.22  & 3.78 $\pm$ 0.06 & 3.90 $\pm$ 0.05 \\
    MAViL \cite{huang2024mavil} & 2 & 12.08 & 12.17 & 2.49 & 1.26 & 2.52 & 90.17 & 3.75 $\pm$ 0.06 & 3.46 $\pm$ 0.07  \\

    \bottomrule 
    \end{tabular}
    }
\end{table}

\begin{table}[t]
    \caption{Ablation results on RFM objective re-weighting.}
    \label{tab:abl_reweight}
    \small
    \centering
    \resizebox{\textwidth}{!}{
    \begin{tabular}{cccccccccc}
    \toprule
    Re-weighting & FD$\downarrow$ & IS$\uparrow$ & KL$\downarrow$ & FAD$\downarrow$ & KID$(10^{-3}) \downarrow$ &  Acc(\%) $\uparrow$ & MOS-Q$\uparrow$ & MOS-A$\uparrow$ \\ 
    \midrule

    \ding{55}  & 11.95 & 12.20 & 2.73 & 1.25 & 2.12 & 97.04  & 3.74 $\pm$ 0.07 & 3.82 $\pm$ 0.06 \\
    \ding{51}  &  12.26 & 12.42 & 2.73 & 1.32 & 2.49 &  97.22 & 3.78 $\pm$ 0.06 & 3.90 $\pm$ 0.05 \\

    \bottomrule 
    \end{tabular}
    }
\end{table}

\subsubsection{Classifier-free guidance scale}

\begin{wrapfigure}{r}{0.65\textwidth}
    \vspace{-1em}
    \includegraphics[width=0.65\textwidth]{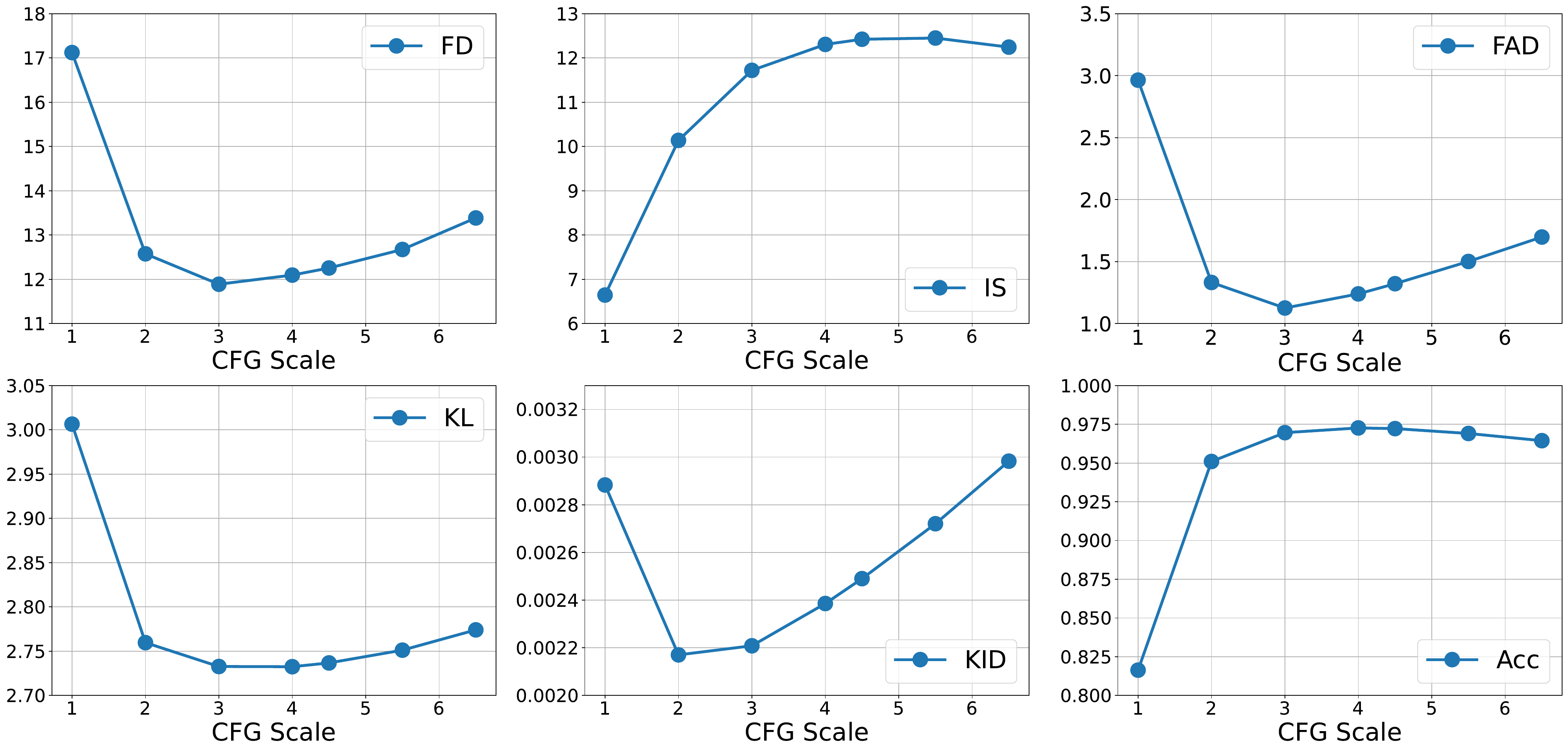}
    \caption{Model performance of \textsc{Frieren} under different CFG scales.}
    \label{fig:cfg}
    \vspace{-1em}
\end{wrapfigure}

In figure \ref{fig:cfg}, we illustrate the impact of various CFG scales on the performance of \textsc{Frieren}. In terms of audio diversity (FD, KL, KID, FAD), the metrics initially increase with the CFG scale, reaching an optimal value at around 2 and 3. After that, the metrics go down as the increasing CFG scale suppresses the diversity. For audio quality (IS) and temporal alignment, as larger scales make the content of the generated audio closer to the visual information, the metrics initially increase with the scale, reaching an optimal value between 4 and 4.5, and decrease after that due to audio distortion. We prioritize audio quality and synchrony and adopt a CFG scale of 4.5.

\subsubsection{Re-weighting RFM objective}

We conduct ablation on RFM objective re-weighting and report the results in table \ref{tab:abl_reweight}. We can see that compared to the vanilla objective, introducing re-weighting results in improvements of 0.22 and 0.18\% for IS and alignment accuracy. This validates the positive impact of objective re-weighting on audio quality and temporal alignment. On the other hand, objective re-weighting causes a decrease in audio diversity, with differences in FD, FAD, and KID being 0.31, 0.07, and 0.37$\times 10^{-3}$, respectively.

\section{Conclusion}

In this paper, we propose \textsc{Frieren}, an efficient video-to-audio generation model based on rectified flow matching. We use a neural network to regress the conditional transport vector field with straight paths from noise to spectrogram latents, and conduct sampling by solving ODE, achieving better performance than diffusion-based and other V2A models. We adopt a vector field estimator based on a feed-forward transformer as well as channel-level cross-modal feature fusion to realize strong audio-video synchrony. Through a combination of reflow and one-step distillation, our model can generate high-quality audio with a few or even one sampling step, boosting the generation efficiency significantly. Experiments show that our model achieves state-of-the-art V2A performance on VGGSound. For future work, we will explore extending the model to larger scales and larger datasets to achieve V2A generation on a broader data domain. Besides, we will attempt audio generation from longer video sequences with variable lengths, rather than being limited to fixed-length short clips. These efforts aim to build a more versatile and widely applicable V2A model.

\section*{Acknowledgment}
This work is supported by the National Natural Science Foundation of China under Grant No. 62222211 and No.62072397.


\bibliographystyle{neurips_2024}
\bibliography{neurips_2024}

\appendix
\clearpage

\section{Implementation details}
\label{appendix:param}

\begin{table}[ht]
    \caption{Architecture details of 1D VAE for spectrogram compression.}
    \label{tab:vae}
    \centering
    \begin{tabular}{lc}
    \toprule
    Hyperparameter & 1D VAE \\
    \midrule
    Input tensor shape for 10-sec audio & (80,624) \\
    Embedding dimension & 20 \\
    Channels & 224 \\
    Channel multiplier & 1, 2, 4 \\
    Downsample layer position &  after block 1 \\
    Attention layer position &  after block 3 \\
    Output tensor shape for 10-sec audio &  (20,312) \\
    \bottomrule
    \end{tabular}%
\end{table}

\begin{table}[ht]
\caption{Hyperparameters of the vector field estimator of \textsc{Frieren} with different sizes.}
\label{tab:hyperparameters}
\centering
\begin{tabular}{lccc}
\toprule
Hyperparameter & Small & Base & Large \\ 
\midrule
Layers &  4 & 4 & 6     \\
Hidden dimension  & 384 & 576 & 768  \\    
Attention heads & 8 & 8 & 8   \\  
Conv1D-FFN dimension  &  1,536 & 2,304 & 3,072   \\    
Number of parameters & 70.90M & 158.88M & 421.12M \\
\bottomrule
\end{tabular}
\end{table}

In table \ref{tab:vae}, we provide the architecture details of the mel-spectrogram VAE. Different from the commonly used 2D VAE for spectrogram, the 1D VAE we adopt does not involve an extra channel dimension, but takes the frequency axis of the spectrogram as the channel dimension, and conducts convolution along the temporal axis. This design is derived from the insight that the spectrogram is not translation invariant along the frequency axis, and it can better synergize with the feed-forward transformer. In table \ref{tab:hyperparameters}, we present the hyperparameters of the vector field estimator networks with different sizes.

Additionally, we observe that although there is no significant difference in objective metrics, initializing the vector field estimator with the weights of a diffusion model for text-to-audio (T2A) generation \cite{huang2023make} helps improve the subjective perceptual quality of the generated audio marginally. This improvement derives from the knowledge of audio generation on a broader data domain learned by the T2A model. We adopt this trick in our model training.

\section{Subjective evaluation}
\label{appendix:eval}

\begin{figure}[h]
\centering
\includegraphics[width=0.8\textwidth]{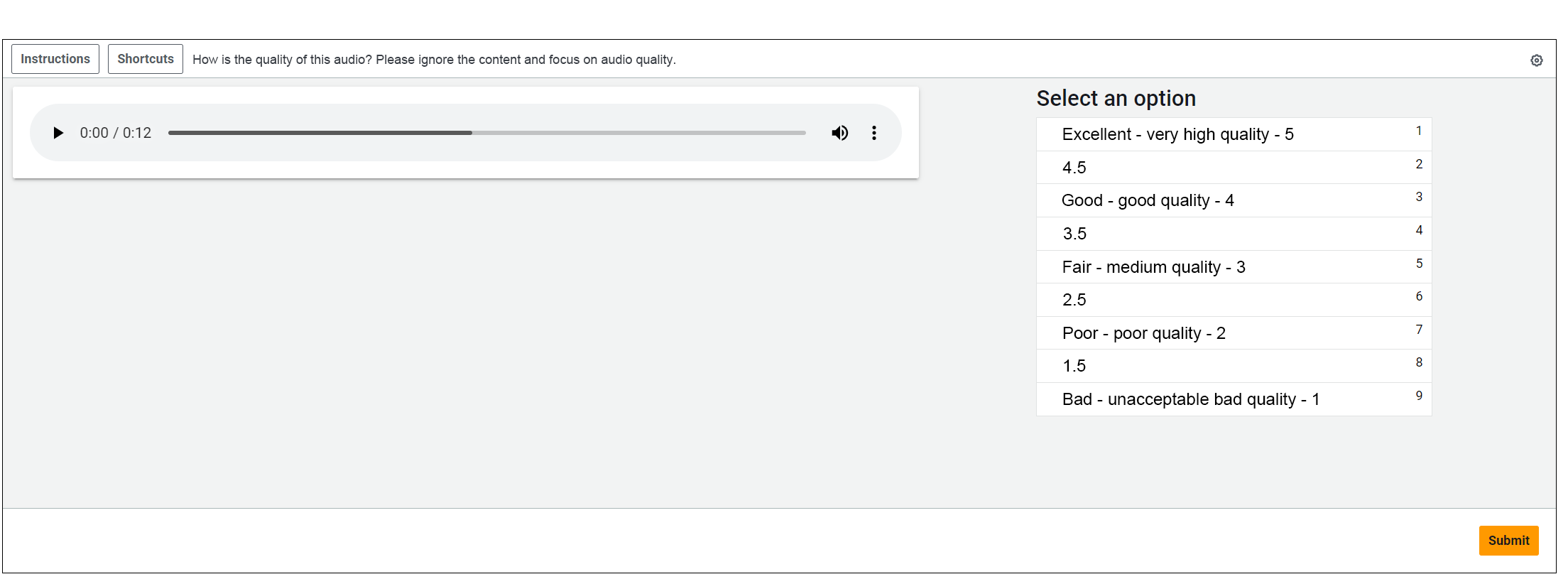}
\caption{Screenshot of subjective evaluation on audio quality.}
\label{fig:mos_q}
\end{figure}

\begin{figure}[h]
\centering
\includegraphics[width=0.8\textwidth]{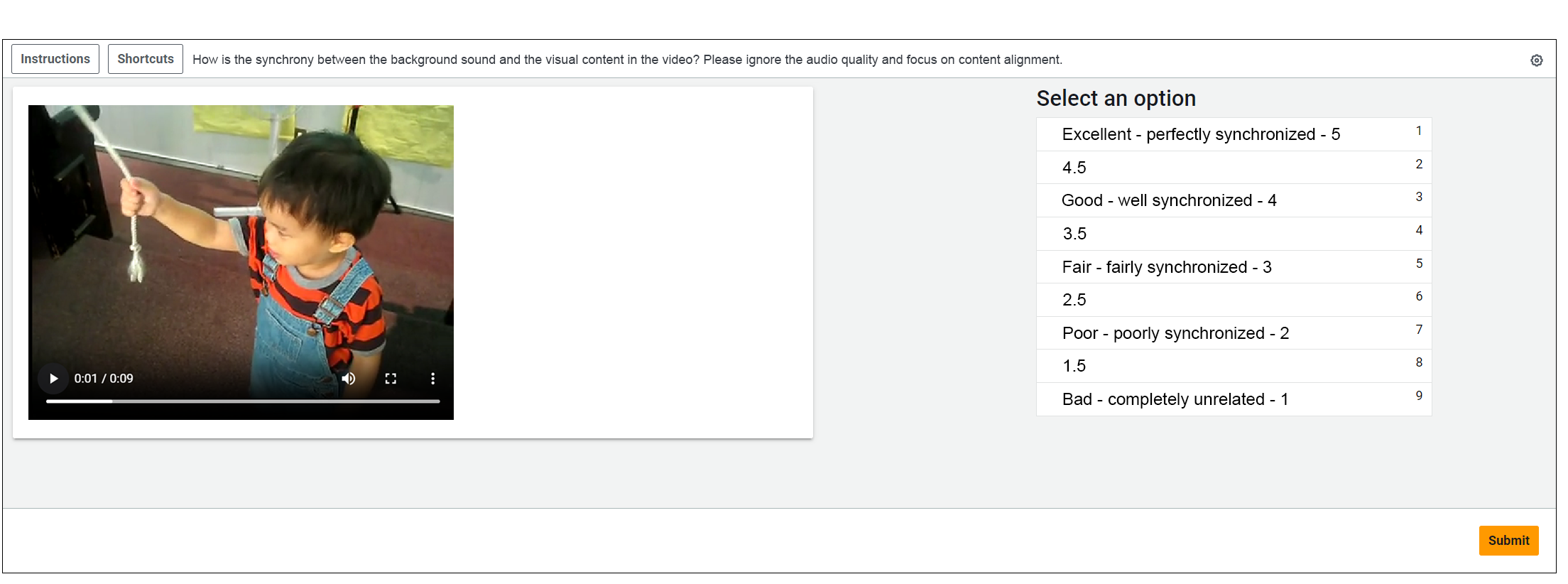}
\caption{Screenshot of subjective evaluation on temporal alignment.}
\label{fig:mos_a}
\end{figure}

For each evaluated model, we select 150 items for subjective evaluation, accounting for about 1\% of the entire test split.

Our subjective evaluation tests are crowd-sourced and conducted via Amazon Mechanical Turk. For audio quality evaluation, we ask the testers to examine the audio quality and ignore the content. And for temporal alignment, we instruct the testers to evaluate the synchrony between the background audio and the video content, while ignoring the audio quality. The testers rate scores on 1-5 Likert scales. We provide screenshots of the testing interfaces in figure \ref{fig:mos_q} and \ref{fig:mos_a}. Each data item is rated by 6 testers, and the testers are paid \$8 hourly.

\section{Time efficiency}
\label{appendix:time}

\begin{table}[h]
    \caption{Inference time per sample of different models with batch size = 1.}
    \label{tab:time}
    \centering
    \begin{tabular}{lc}
    \toprule
    Model & Inference Time (sec) \\
    \midrule
    SpecVQGAN & 3.936 \\
    Im2Wav & 333.246 \\
    Diff-Foley  (step=25) & 2.104  \\
    \textsc{Frieren} (Dopri5, step=25) & 1.510 \\
    \textsc{Frieren} (Euler, step=25) & 0.288 \\
    \textsc{Frieren} (Euler, step=5) & 0.064 \\
    \textsc{Frieren} (Euler, step=1) & 0.031 \\
    \bottomrule
    \end{tabular}%
\end{table}

In table \ref{tab:time}, we compare the inference time per sample of different models. The inference is conducted on a single RTX-4090 GPU with a batch size of 1. We can see that the inference procedure of transformer-based autoregressive models, including SpecVQGAN and Im2Wav, is more time-consuming, especially for Im2Wav, which takes several minutes to generate a single sample. This is because Im2Wav conducts a cascaded generation with 2 transformers. Moreover, its use of high-bitrate audio VQVAE results in very long sequences of audio representation, significantly increasing the inference time required for the transformers, which has quadratic time complexity concerning sequence length. In contrast, Diff-Foley and \textsc{Frieren} require less inference time, and \textsc{Frieren} with Euler solver enjoys a higher speed, achieving 7.3 times faster than Diff-Foley with 25 sampling steps. 
This is the result of a combination of multiple factors, including model architecture, model parameters, sampling methods, and so on. Furthermore, when using \textsc{Frieren} model with reflow and one-step distillation, we can generate 5-step sampled audio in 0.064 seconds and 1-step sampled audio in just 0.031 seconds, achieving 4.5$\times$ and 9.3$\times$ acceleration compared to 25-step sampling. This demonstrates the extremely high generation efficiency of our model on the task of V2A.

\section{Impact of the vocoder on model performance}

\begin{table}[t]
    \scriptsize
    \caption{Comparison of the performance of Diff-Foley and \textsc{Frieren} using the same vocoder.}
    \label{tab:res_voc}
    \footnotesize
    \centering
    \begin{tabular}{lccccc}
    \toprule
    Model & FD$\downarrow$ & IS$\uparrow$ & KL$\downarrow$ & FAD$\downarrow$ & KID$(10^{-3}) \downarrow$ \\ 
    \midrule
    \multicolumn{6}{l}{BigVGAN} \\
    \midrule
    Diff-Foley (CG \ding{51}) &  18.02 & 10.89 & 2.88 & 6.32 & 5.32  \\
    \textsc{Frieren} & \textbf{12.26} & \textbf{12.42} & \textbf{2.73} & \textbf{1.32} & \textbf{2.49}  \\
    \midrule
    \multicolumn{6}{l}{Griffin-Lim} \\
    \midrule
    Diff-Foley (CG \ding{51}) & \textbf{23.94} & \textbf{11.11} & 3.38 & 4.72 & \textbf{9.58}  \\
    \textsc{Frieren} & 28.29 & 10.67 & \textbf{3.17} & \textbf{3.70} & 12.30  \\
    \bottomrule 
    \end{tabular}

\end{table}

Different selections of vocoders can significantly impact the performance of various audio generation models. Diff-Foley uses the simple Griffin-Lim method \cite{griffin1984signal} to map spectrograms to waveforms, while \textsc{Frieren} employs the more efficient BigVGAN. To compare the performance of the spectrogram generation models while minimizing the influence of the vocoder, we apply BigVGAN and Griffin-Lim separately to each model. The output from Diff-Foley is converted into an 80-bin mel-spectrogram and then fed into BigVGAN. The number of Griffin-Lim iterations for \textsc{Frieren} is the same as Diff-Foley. The results are shown in table~\ref{tab:res_voc}.

It can be seen that using BigVGAN for Diff-Foley improves its FD, KL, and KID, indicating its effectiveness. On this basis, \textsc{Frieren} outperforms Diff-Foley across all metrics, with a greater difference than when using Griffin-Lim for both. This further demonstrates that our model is superior to Diff-Foley. 

On the other hand, when using Griffin-Lim for both models, despite the performance drop, \textsc{Frieren} still surpasses Diff-Foley in KL and FAD, with FAD showing a significant advantage while maintaining competitive FD and IS values. We speculate that the Griffin-Lim algorithm is so weak that it forms a performance bottleneck, narrowing the performance gap between \textsc{Frieren} and Diff-Foley. Additionally, differences in spectrogram hyperparameters may also lead to a performance gap. Diff-Foley uses 128 frequency bins, more than the 80 bins used by \textsc{Frieren}, allowing it to carry finer-grained information and may give Diff-Foley an advantage when using Griffin-Lim.

\section{Limitations and boarder impacts}
\label{appendix:limit}

\paragraph{Limitations} Despite that \textsc{Frieren} achieves outstanding performance on audio quality, temporal alignment, and generation efficiency, it still has two major limitations: 1) Currently, experiments have only been conducted on a small-scale dataset, VGGSound, and we have not yet scaled the model to large-scale datasets. Therefore, it is still difficult to apply our model to a wide range of real-world scenarios for now; 2) our current model design only targets audio generation for fixed-length short video clips, and it lacks the ability of audio generation for long videos with various lengths. We will explore the solutions to these issues in future work.

\paragraph{Potential positive impacts} The achievements of our model on the V2A task may reduce the cost of sound effect synthesis, and could potentially drive advancements in the film, gaming, and social media industries. 

\paragraph{Potential negative social impacts} The automatic sound effect generation technology may lead to job losses for related personnel. Additionally, there is a risk of the model being used to generate harmful content or fake media. Constraints are needed to guarantee that people will not use the model in illegal cases.

\end{document}